\documentclass[twoside]{article}

\font\ff=cmss12 \font\ft=cmbx12

\def\farcs{\hbox{$.\!\!^{\prime\prime}$}}  
\def\fmm{\hbox{$.\!\!^{\rm m}$}}           

\input epsf.sty

\begin{document}
\thispagestyle{empty}
  \centerline {\ff COMMISSIONS 27 AND 42 OF THE IAU}
  \centerline {\ff INFORMATION BULLETIN ON VARIABLE STARS}
  \vskip 0.1cm
  \centerline {\rm Number }
  \vskip 0.5cm
  \hbox to \textwidth{\hfil
  {\vbox{\hbox{\rm Konkoly Observatory}\hbox{\rm Budapest}
  \hbox{\rm  }\vskip 0.2cm \hbox{\sl HU ISSN 0374 -- 0676}}}}
  \vskip 0.8cm plus 0.2cm minus 0.4cm
%
{\uppercase
  {\centerline {\ft Variability of V838 M\lowercase{on} Before Its
Outburst} \vskip 0.5cm plus 0.4cm minus 0.2cm}}

{\uppercase{ Kimeswenger, S.$^1$; Eyres, S.P.S.$^2$} \vskip 0.3cm
plus 0.1cm minus 0.1cm}

{\small $^1$Institut f{\"u}r Astro- und Teilchenphysik,
Universit{\"a}t Innsbruck,
      Technikerstra{\ss}e 25, A-6020 Innsbruck,
      Austria\protect\newline
\indent$^2$Dept. of Physics, Astronomy \& Mathematics, University
of Central Lancashire, Preston PR1 2HE, UK
        \vskip 1mm \large}

\vskip 1cm plus 0.4cm minus 0.4cm
        \large \baselineskip=5mm
        \markboth{{\ff IBVS}  }{{\ff IBVS}  }
        \renewcommand{\thefootnote}{\arabic{footnote}}
        \setcounter{footnote}{0}
        \noindent

V838 Mon had an unusual "nova-like" outburst in 2002 (Munari et
al. 2002, Kimes\-wenger et al 2002). Several attempts at
photometry of the progenitor on archival plates led to different
results (Munari et al. 2002 Kimeswenger et al. 2002, Goranskij et
al. 2004). While the first two used scans based on the SERC J
plate from 1983 and the UKST ER plate from 1989, Goranskij et al.
(2004) added the UKST I plate from 1982 and the POSS-I plates from
1952. Munari et al. (2005) used the USNO-B1 catalogue and a
revised calibration based on their CCD sequence. The USNO-B1 is
based on scans with an 8 bit linear greyscale only. Thus the
stellar images are without grey wings and no de-blending can be
done. There are also two bright objects (USNO-B1.0 0861-0120005
and USNO-B1.0 0861-0120000) at/near the target position. It is
also not clear to the reader how Munari et al. (2005) averaged the
different bands used in USNO-B1 (POSS-I O and SERC-J). The
investigation of the "older twin" of this unusual object - V4332
Sgr (Nova Sgr 1994) - shows the progenitor might be variable
during the last years before outburst (Kimeswenger, 2006). This is
essential for the investigation of the spectral energy
distribution (SED).

\begin{figure}[ht]
\centerline{\epsfysize=6cm\epsfbox{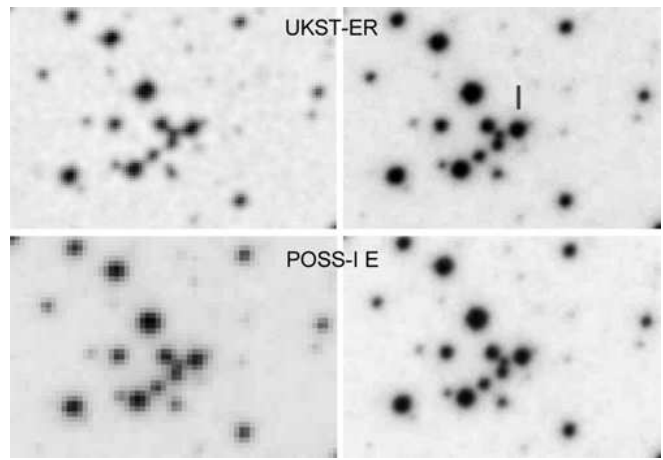}}\caption{The
upper panel shows the 1989 UKST-ER plate. The left hand image is
the DSS-2 scan (resolution 1\farcs01). The right image shows the
same plate from the SuperCOSMOS scans (resolution 0\farcs67). On
the DSS-2 scan the stars are clearly elongated and overlap with
their neighbors. The lower panel shows the same field on the DSS-1
(resolution 1\farcs7) POSS-I E plate and again the SuperCOSMOS
scan. Here the de-blending problem for V838 Mon is even more
obvious.}
\end{figure}

Here we used not DSS scans, but the SuperCOSMOS scans (except
POSS-I O) of the same plates used by Munari et al. (2002) and
Goranskij et al. (2004). These scans have a much higher spatial
resolution. Bacher et al. (2005) have shown, that this does not
normally improve the photometry of unblended stars. But as already
mentioned there, blended objects have often been rejected in their
work. V838 Mon is within a small group of stars. Small apertures
and high resolution are thus essential here (for a calibration of
the "best aperture" see Bacher et al. 2005). Figure 1 shows the
increase of quality and better de-blending capabilities using the
SuperCOSMOS scans. In addition to the surveys used up to now, the
SuperCOSMOS server also provides us with the scans of the new
UKST-SR survey. This plate was taken in parallel to the UKST
H$\alpha$ survey for off-band continuum subtraction. It was
obtained less than four years before the outburst of V838 Mon and
was overlooked up to now. It gives us valuable additional
information. All photographic plates were calibrated using the CCD
sequence of Munari et al. (2005) and the nonlinear tuning for
digitized sky surveys by Bacher et al. (2005). The latter change
of the method is the main difference to the calibration used by
Kimeswenger et al. (2002). They only used a linear approximation
to a few stars having the same magnitude.

\begin{figure}[ht]
{\epsfxsize=\textwidth\epsfbox{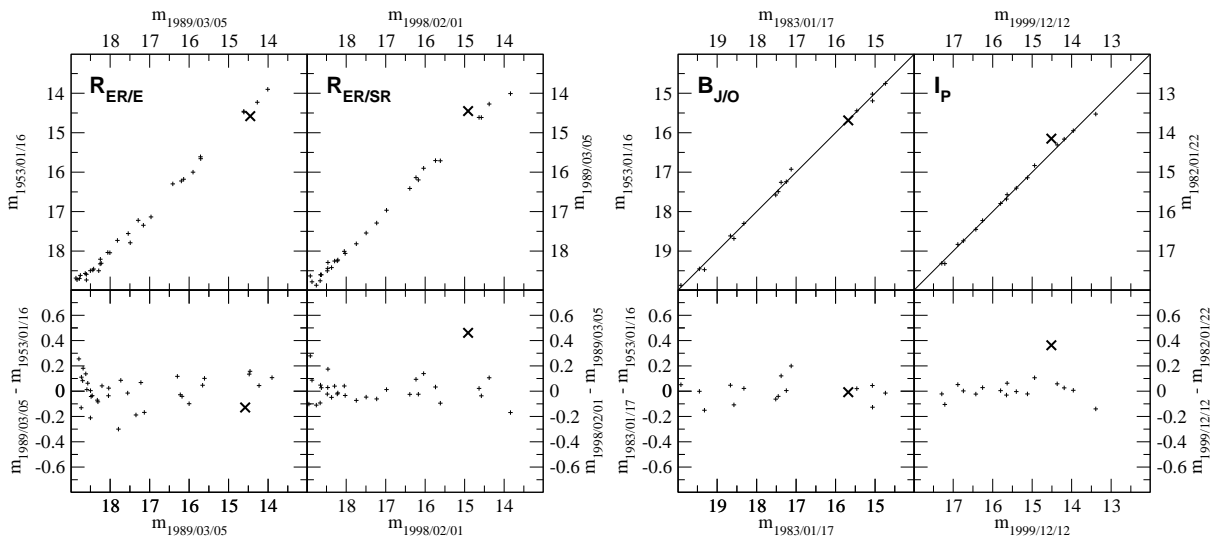}}\caption{The
photographic red band photometries from POSS-I E (1953) and
UKST-ER (1989) show no significant variation until at least 1989.
The fading of V838 Mon (cross) during the late nineties is evident
on the UKST-SR (1998) plate. The blue photometries (converted to
standard B magnitudes) show no variations before 1983 either. The
I band combines the photographic UKST-IR (1982) plate with the
data from the DENIS (1999) CCD survey. The same fading as in the
R-band is obvious.}
\end{figure}
The blue bands of the POSS-I and of  the SERC-J survey strongly
differ in their bandpass. Thus the conversion to standard B
magnitudes was used for the comparison. While the B$_{\rm J}$
conversion is well studied (Bacher et al. 2005) there exist no
such extensive studies for the POSS-I O. Dorschner et al. (1966)
assume there is no color correction required. We found with the
field stars $m_{\rm O} = {\rm B} - 0.030 ({\rm B-R}) - 0.058$.
This correction was applied to derive $m_{\rm O}$ magnitudes of
the stars of the CCD sequence for calibration purposes. As most of
the field stars are foreground stars with typically 0\fmm4 $\le$
(B-V) $\le$ 0\fmm8, this effect is small. This led Goranskij et
al. (2004) to the conclusion, that color corrections need not be
applied at all. They used a comparison with stars in that color
range only. While these field stars do have spectral types of A-F
with a strong Balmer jump, the progenitor of V838 Mon is a heavily
reddened blue object without any Balmer jump. Thus the effective
wavelength differs even when they have about the same (B-V) color.
This is certainly true for the B$_{\rm J}\rightarrow$B conversion.
However it is weak at the wavelengths of the SERC-J survey, so it
may not affect the work of Goranskij et al. (2004) significantly.
It is more significant for the $m_{\rm O}\rightarrow$B calculation
(with the filter just on the Balmer continuum absorption).

The last data before the outburst was taken by the DENIS and the
2MASS surveys. The 2MASS survey visited the target twice due to an
overlap of neighboring tiles. While the 02/11/1998 data is in the
point source data base, another plate was taken just 37 days after
that. We have loaded both images from the data base, to redo the
photometry on both of them. This gives a good error estimate by
using the stars in the overlap of the two observations. Finally we
have access to the non--public DENIS images. The DENIS survey is
known to sometimes have systematic zero point shifts. The standard
survey operations calibration is insufficient here. Also the
K$_{\rm s}$ band was at its limits for this source. Thus the
question arises, whether the difference to 2MASS is therefore high
in this band. Using the 2MASS data of the field stars around the
target and the improvement of the calibration for DENIS data by
Kimeswenger et al. (2004) We obtained a more accurate photometric
calibration in J and K$_{\rm s}$. The corrected values are given
in the table below.
\begin{figure}[ht]
\centerline{\epsfysize=6cm\epsfbox{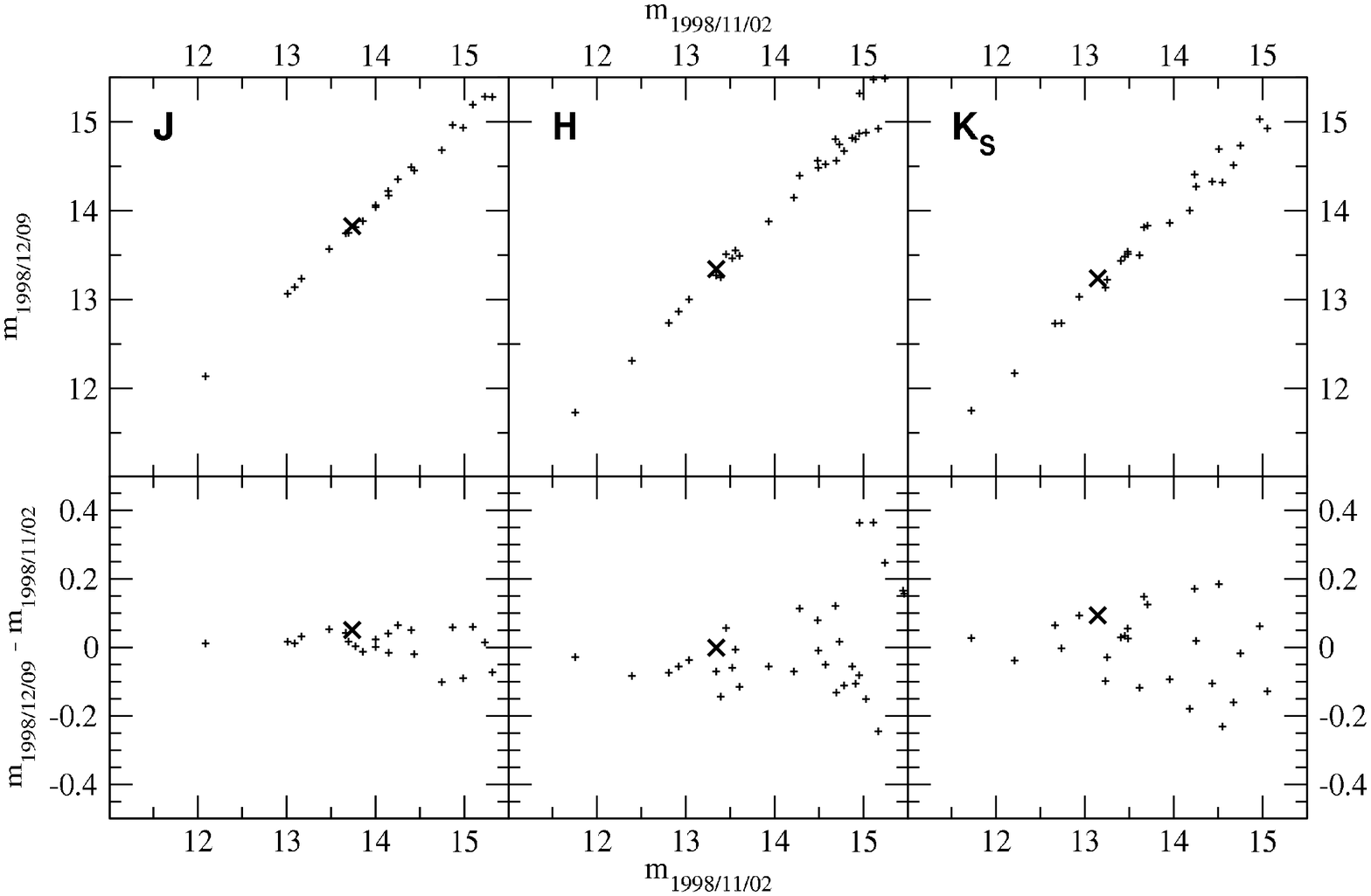}}\caption{The
2MASS data obtained 2/11/98 vs. those taken 9/12/98.}
\end{figure} 
The target seems to be stable before 1990. This corresponds to the
finding of Goranskij et al (2004) who had their last Sonneberg
plate 13/03/1991. After this a fading clearly started. The 2MASS
data gives a weak indication in all three bands, that this fading
continued in late 1998. At the end of 1999 the DENIS J and K$_{\rm
s}$ data show a small re-brightening by about 10\%. This is also
consistent with the fact that I$_{1999}-$I$_{1982}$ = 0\fmm363 is
different from R$_{1998}-$R$_{1989}$ = 0\fmm461 by about
K$_{1999}-$K$_{1998}$ $\simeq$ J$_{1999}-$J$_{1998}$ $\simeq$
-0\fmm1 .


\begin{table}[ht] \caption{Summary of the
photometry (sorted by date of observation). The horizontal line in
the middle marks the start of the fading. Data before this line
should not be mixed with those after  the line, when adjusting a
SED.}
\begin{center}
\begin{tabular}{l l l c c c l}
 \noalign{\smallskip}
\hline
\hline
date & JD - & material & band name & ${\lambda_{eff}}^{1)}$  & mag & err\\
  & 2400000.0 & & & [$\mu$m] & &  \\
\hline
16/01/1953 & 34393.32 & POSS-I    & E           & 0.650 & 14\fmm58 & 0\fmm13 \\
16/01/1953 & 34393.41 & POSS-I    & O           & 0.405 & 15\fmm68 & 0\fmm15 \\
22/01/1982 & 44990.57 & SERC-I    & I$_{\rm p}$ & 0.840 & 14\fmm15 & 0\fmm08 \\
17/01/1983 & 45350.52 & SERC-J    & B$_{\rm J}$ & 0.475 & 15\fmm49 & 0\fmm09 \\
05/03/1989 & 47589.47 & UKST-ER   & r           & 0.650 & 14\fmm45 & 0\fmm09 \\
\hline
01/02/1998 & 50844.45 & UKST-SR   & r           & 0.650 & 14\fmm91 & 0\fmm10 \\
02/11/1998 & 51119.86 & 2MASS     & J           & 1.150 & 13\fmm86 & 0\fmm03 \\
           &          &           & H           & 1.650 & 13\fmm50 & 0\fmm04 \\
           &          &           & K$_{\rm s}$ & 2.150 & 13\fmm31 & 0\fmm04 \\
09/12/1998 & 51156.83 & 2MASS     & J           & 1.150 & 13\fmm96 & 0\fmm04 \\
           &          &           & H           & 1.650 & 13\fmm55 & 0\fmm03 \\
           &          &           & K$_{\rm s}$ & 2.150 & 13\fmm43 & 0\fmm05 \\
12/12/1999 & 51524.76 & DENIS     & I$_{\rm c}$ & 0.790 & 14\fmm52 & 0\fmm03$^{2)}$ \\
           &          &           & J           & 1.150 & 13\fmm82 & 0\fmm06 \\
           &          &           & K$_{\rm s}$ & 2.150 & 13\fmm12 & 0\fmm07 \\
\hline \hline
\end{tabular}\newline
{\footnotesize 1) based the the SED with T$_{eff} > 15\,000\,$K
and E(B-V) $\approx$ 0\fmm7\phantom{MMMMMMMMMMMMMMMMMMMMMMMM}
\newline\vspace{-0.5cm}

 2) single band - thus error estimate taken from
survey point source catalogue\phantom{MMMMMMMMMMMMMMMMMMn}}
\end{center}
\end{table}

In our opinion the  discrepancies of the photometry mentioned in
the introduction originate in the blend with neighboring objects
and the different handling of color equations. The new photometry
provided here now gives more accurate values for SED fitting. The
fading found here might be important for interpreting the nature
of this unique object. But even more important is the fact that
the photographic data before 1990 should not be used together with
the 1998/1999 NIR survey data when fitting the SED or when
deriving the foreground extinction. The fading lowered the NIR
data and thus leads to an overestimate of the interstellar
extinction and/or an overestimate of the progenitors effective
temperature. As we do not have blue data during the late nineties,
we do not have any idea about a possible color change. Thus we
cannot decide, if the fading is caused by a change of the
temperature, a contraction of the photosphere, or any other kind
of geometric effects.


\newpage

{\vskip 1cm plus 0.5cm minus 0.5cm \noindent Reference: \hfil
  \vskip 0.3 cm plus 0.1 cm minus 0.1cm
\parindent=-1cm \leftskip 1cm}

Bacher A., Kimeswenger S., Teutsch P., 2005, {\it MNRAS}, {\bf 362}, 542


Dorschner J., G\"urtler J., Schielicke R., Schmidt K.-H., 1966,
{\it AN}, {\bf 289}, 51

Goranskij V.P., Shugarov S.Yu., Barsukova E.A., Kroll P., 2004, {\it IBVS}, {\bf 5511}

Kimeswenger S., 2006, {\it AN}, {\bf 327}, 44

Kimeswenger S., Lederle C., Richich A., et al., 2004, {\it A\&A}, {\bf 413}, 1037

Kimeswenger S., Lederle C., Schmeja S., Armsdorfer B., 2002, {\it
MNRAS}, {\bf 336}, L43

Monet D.G., Levine S.E., Canzian B., et al., 2003, {\it AJ}, {\bf
125}, 984

Munari U., Henden A., Kiyota S., et al., 2002, {\it A\&A}, {\bf
389}, L51

Munari U., Henden A., Vallenari A, et al., 2005, {\it A\&A}, {\bf 434}, 1107

{\vskip 1cm plus 0.5cm minus 0.5cm
\parindent=15.0pt \leftskip -0cm}

\end{document}